\documentclass[aip,amsmath,amssymb,reprint]{revtex4-1}
\usepackage{graphicx}% Include figure files
\usepackage{dcolumn}% Align table columns on decimal point
\usepackage{bm}% bold math

\begin{document}

\title{Improving Li$_2$O$_2$ conductance via polaron preemption: an \textit{ab initio} study of Si doping}% Force line breaks with \\

\author{V. Timoshevskii}
  \email{Timochevski.Vladimir@ireq.ca}
  \affiliation{Institut de recherche d'Hydro-Qu\'ebec (IREQ), 1800, boul. Lionel-Boulet, Varennes (Qu\'ebec),
Canada J3X 1S1}

\author{Zimin Feng}
  \affiliation{Institut de recherche d'Hydro-Qu\'ebec (IREQ), 1800, boul. Lionel-Boulet, Varennes (Qu\'ebec),
Canada J3X 1S1}
  \affiliation{D\'epartement de g\'enie des mines et mat\'eriaux, Division de g\'enie des mat\'eriaux, McGill University, Montr\'eal (Qu\'ebec), 
Canada H3A 0C5}

\author{K. Bevan}
 \affiliation{D\'epartement de g\'enie des mines et mat\'eriaux, Division de g\'enie des mat\'eriaux, McGill University, Montr\'eal (Qu\'ebec), 
Canada H3A 0C5}

\author{K. Zaghib}
\affiliation{Institut de recherche d'Hydro-Qu\'ebec (IREQ), 1800, boul. Lionel-Boulet, Varennes (Qu\'ebec),
Canada J3X 1S1}

%% \date{\today}% It is always \today, today,

\begin{abstract}
We report on \textit{ab initio} electronic structure simulations of Li$_2$O$_2$, where 1.6\% of lithium atoms are substituted by silicon. It is demonstrated that this leads to the formation of conducting impurity states in the band gap of Li$_2$O$_2$. We show that these states originate from the antibonding orbitals of the oxygen pairs and are remarkably stable against possible polaron formation (upon electron injection).  Through this polaron preemption mechanism, the proposed compound is expected to show significantly higher electronic mobility than stoichiometric Li$_2$O$_2$, which could have significant applications in lithium-air batteries.
\end{abstract}

%%% \pacs{Valid PACS appear here}

\maketitle

% main text
%%% \section{Introduction}

In recent years lithium-air (Li-air) batteries have received considerable attention from both experimentalists and theoreticians, due to their exceptionally high theoretical energy density.\cite{Armand08, Abraham96} However, their practical implementation has been severely stalled by a series of technical challenges, including: poor cyclability, low electrical efficiency, and reduced capacity at high discharge rates.\cite{Yang10} These obstacles though to partially arise from strong polarization losses at the cathode material; which, in turn, results from the poor conductivity of lithium peroxide (Li$_{2}$O$_{2}$) -- the main discharge product, forming at the cathode of Li-air batteries.\cite{Kang12, Hummelshoj10}

Recent theoretical studies have demonstrated that stoichiometric Li$_{2}$O$_{2}$ is a wide band gap insulator, and that charge transport in this compound takes place via polaron hopping.\cite{Hummelshoj10, Kang12, Ong12} In their first-principles theoretical study, Kang \textit{et al} \cite{Kang12} demonstrated that an electron injected into Li$_{2}$O$_{2}$ will be self-trapped through the formation of a small polaron.  It was further shown that polaron formation arises due to the molecular nature of the Li$_{2}$O$_{2}$ conduction band, whereby an injected electron is localized at an O-O site possessing an elongated oxygen pair. The calculated hopping rate of this small polaron was directly correlated to the extreemly low electronic mobility of Li$_{2}$O$_{2}$. At the same time, Hummelsh{\o}j \textit{et al} \cite{Hummelshoj10} argued that Li vacancies, which are likely to be created during the discharge process, may induce delocalized hole states in the valence band of lithium peroxide, therefore improving Li$_{2}$O$_{2}$ charge transport via hole conduction. However, Ong \textit{et al} \cite{Ong12} demonstrated that hole states may also become localized through the formation of ``hole'' polarons.  Although the migration barriers of hole polarons were determined to be lower than electronic polarons, the resulting conductivity of Li$_{2}$O$_{2}$ was still far below technologically desired values.  Thus far it seems the only practically feasible way of improving lithium peroxide's conductivity, proposed up to the present moment, relies on the use of ``contact states'' : Zhao \textit{et al} 
\cite{Zhao12} demonstrated that growing of Li$_{2}$O$_{2}$ on a graphene layer induces the hole-type conducting channels in lithium peroxide. However,  this approach is limited by the short penetration depth of surface states into Li$_{2}$O$_{2}$ layers.    Such states only penetrate only a few LiO$_2$ layers inside the lithium peroxide crystal, thus limiting conduction enhancement to only a thin surface layer.

In this Letter we propose an alternative way of improving the conductivity of lithium peroxide. By means of \textit{ab initio} calculations, we show that the substitution of a small fraction (1.6\%) of Li atoms by Si impurities leads to the creation of additional conducting states in Li$_{2}$O$_{2}$. These states are shown to originate from the partial occupation of oxygen antibonding orbitals by electrons, donated by impurity Si atoms. The elongated oxygen pairs, forming these states, are bound to Si impurities, and are not subject to polaron-induced deformations.   This polaron preemption mechanisim is predicted to significantly enhance the electron mobility of lithium peroxide.

%%%% \section{Computation details}

Our study was performed within density functional theory (DFT).\cite{HK-KS}   Exchange-correlation interactions were captured within the generalized gradient approximation (GGA).\cite{PBE}  We utilized the SIESTA package, \cite{Siesta} which employs norm-conserving pseudopotentials and numerical atomic orbitals (NAOs). A flexible basis set consisting of doubled \{$s$\}, tripled \{$p_x,p_y,p_z$\}, and polarized \{$d$\} orbitals was determined essential to obtain a correct description of the electronic structure and energetics of this system.  In all calculations, structures were relaxed both with respect to atomic positions and unit cell size.  Benchmark calculations on stoichiometric Li$_2$O$_2$ produced lattice parameters of $a=3.19$ and $c=7.71~\AA$.  This is in excellent agreement with previous all-electron calculations ($3.18$ and $7.73~\AA$),\cite{Cota05} performed using the same GGA functional, and also agrees well with experimental data ($3.14$ and $7.65~\AA$).\cite{Foppl57} 

We constructed a $4 \times 4 \times 2$ supercell to model silicon impurities in Li$_2$O$_2$. This relatively large 256-atom supercell allowed us to study a system with 2 silicon atoms per unit cell, and to calculate several possible Si distribution patterns (discussed below). We found it necessary to use a  $4 \times 4 \times 4$ \textit{k}-point mesh to ensure total energy convergence. As additional benchmark, we calculated the formation energies of Li vacancies in Li$_2$O$_2$. Using the bulk Li chemical potential as a reference, we obtained the formation energies of 3.39 and 3.52 eV for intralayer and interlayer Li vacancies, respectively. These results are in line with a previous value of 2.8 eV (obtained using the GGA functional\cite{Hummelshoj10}), as well as more recent results of 3.8 and 4.1 eV (calculated using the HSE functional \cite{Ong12}).          

%%% \section{Results and discussion}
We acknowledge that the impurity atom (Si) distribution within a Li$_2$O$_2$ host matrix is a complex problem, that deserves a separate investigation. Many different configurations may exist, including possible dimerization, clusterization, and phase segregation. The invesigation of this complex problem, which should certainly be useful and interesting, is, however, outside of the scope of the present study. The purpose of our investigation is to study the influence of Si impurity atoms on the electronic properties of Li$_2$O$_2$, and to show how these impurities may improve the conductivity of lithium peroxide. Therefore, we have limited our study to several possible impurity distrubution patterns, which were chosen using the following assumptions: (i) we considered an ``impurity regime'', with no direct interaction between Si atoms, by placing them at significant distances from each other; (ii) only substitutional Si doping in the LiO$_2$ layer was considered (intralayer doping), as this configuration was found to be more energetically favorable than the interlayer doping configuration (see below); (iii) keeping the Si-Si distance as large as possible, three patterns of Si distribution in the supercell were tested, including one uniform distribution, and two patterns where Si atoms order along certain directions in the Li$_2$O$_2$ structure.          

\begin{figure}
  \includegraphics[width=8cm]{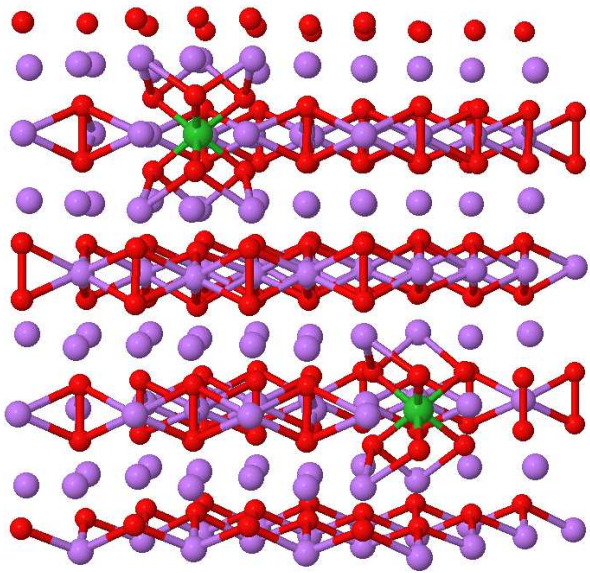}\\
  \caption{(Color online) The $4 \times 4 \times 2$ supercell of Li$_2$O$_2$, used in doping calculations. Li, O, and Si atoms are shown in blue, red, and green colors, respectively. The positions of Si atoms correspond to a uniform impurity distribution, used in the electronic structure analysis.}\label{cell}
\end{figure}

%STOPPED HERE

The 256-atom supercell used in our calculations is shown in Fig.~\ref{cell}.  Within this structure, Si can dope substitutionally at two symmetrically distinct sites: within the LiO$_2$ layer (intralayer doping) and between LiO$_2$ layers (interlayer doping).  We have found the intralayer doping configuration to be 1.1 eV lower in total energy than interlayer doping. To further investigate intralayer doping, three different patterns of widely separated Si dopants were considered in our supercell.  In the first distribution pattern (shown in Fig.~\ref{cell}), Si atoms were placed in two different non-adjacent LiO$_2$ layers along the diagonal of the unit cell. This pattern models a near uniform impurity distribution, with a Si-Si separation distance of 11\AA. In the the second and third patterns (not shown in Fig.~\ref{cell}), we considered two possible orderings of Si impurities along a given direction in the Li$_2$O$_2$ lattice: inside the LiO$_2$ layer and normal to the LiO$_2$ layer. 

Amongst all three structures, the uniform Si distribution (Fig. \ref{cell}) was found to be the most energetically favorable one. The next lowest-energy structure (with in-plane Si ordering) was found to be 0.23 eV higher in total energy. Therefore, only the uniform Si distribution in Fig.~\ref{cell} was considered in followup  calculations. Lastly, we estimated the energetical stability of the Si-doped Li$_2$O$_2$ structure, shown in Fig.\ref{cell}, by calculating its formation energy. Taking as references the chemical potentials of Li and Si in their bulk phases, we obtained the formation energy of the silicon substitutional defect to be -4.9 eV (per Si atom). This demonstrates that the occupation of Li intralayer sites by Si atoms should be energetically favorable during the discharge process.     

\begin{figure}
 \includegraphics[width=8cm]{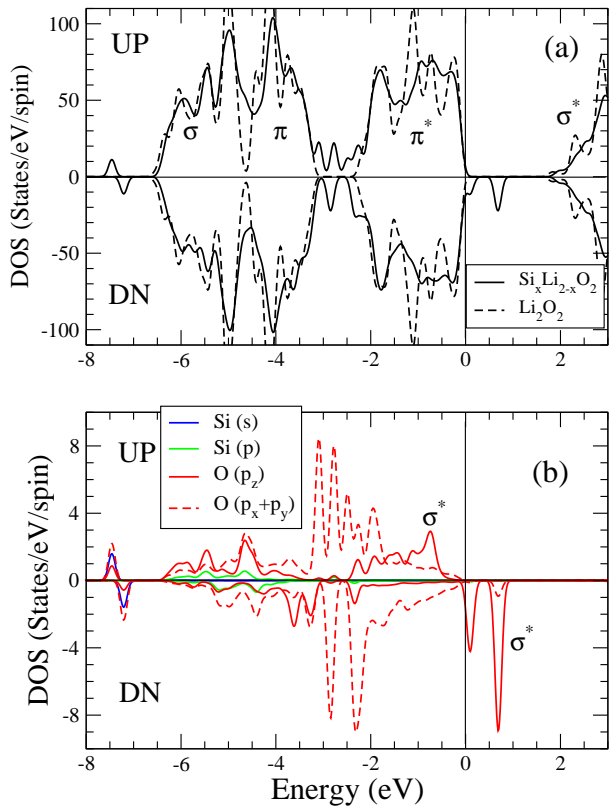}\\
   \caption{Panel (a) shows density of electronic states (DOS) for Si-doped Li$_2$O$_2$ 256-atom supercell. DOS for undoped supercell is also shown for comparison. Panel (b) shows DOS, projected on the atomic orbitals of silicon and on the neighboring three elongated O-O pairs. }\label{dos}
\end{figure} 

Next, we investigated the influence of Si-doping on the electronic structure of lithium peroxide.  The total density of electronic states (DOS) for the doped system is plotted in Fig.~\ref{dos}(a) (solid black line). The DOS for undoped Li$_2$O$_2$ is also shown for direct comparison (dashed black line). The electronic structure of pure lithium peroxide has been extensively studied in literature,\cite{Kang12, Hummelshoj10, Ong12, Zhao12} and our results for the undoped case agree well with previous investigations. The valence and conduction bands of Li$_2$O$_2$ are almost exclusivlely formed by the $2p$-states of the O$_2^{2-}$ oxygen pair. The valence band is composed of $\sigma$, $\pi$, and $\pi^*$ states, while the conduction band has a $\sigma^*$ character. Fig.~\ref{dos}(a) predicts the following main changes in lithium peroxide DOS upon Si doping: (i) a new group of localized states is created at -7 eV near the bottom of the valence band; (ii) additional states appear in the gap between $\pi$ and $\pi^*$ states in the middle of the valence band; (iii) new conducting states appear in the band gap of the system right above the valence band maximum. 

\begin{figure}
 \includegraphics[width=8cm]{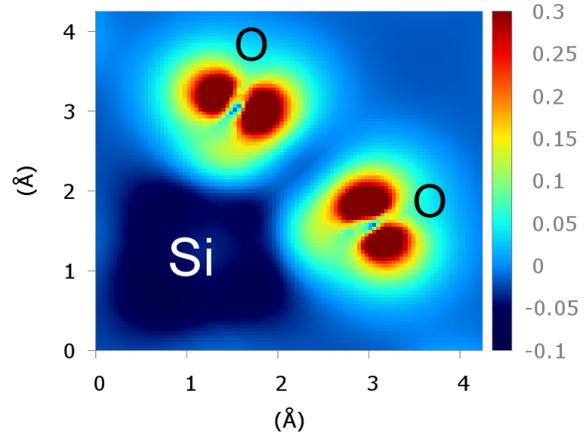}\\
  \caption{The difference between self-consistent charge density and the superposition of atomic charge densities, calculated in the plane, crossing one Si atom and one neighboring O-pair. The charge density is in $e/\AA^3$.}\label{drho}
\end{figure} 

To understand these electronic structure changes, let us begin by investigating the structural changes induced by Si substitution. It can be seen in Fig.~\ref{cell} that Si enters into a bonding interaction with 3 nearby oxygen pairs. This interaction results in the elongation of each O$_2$ pair and the formation of a SiO$_6$ cluster. Interestingly, the O-O distances in this cluster are 2.11~\AA, which is significantly larger than the inter-oxygen distance of unperturbed O$_2$ ions in the system (1.55~\AA), and is nearly identical to the O-O distance of the polaronic site (2.1~\AA) which is formed when extra electron is injected into Li$_2$O$_2$.\cite{Kang12} To trace the origin of this similarity we calculated the density of states of the SiO$_6$ dopant cluster (see Fig. \ref{dos}b). Two immediate observations can be extracted from this plot: (i) the states at $-7$ eV are mostly created through the hybridization of Si $3s$-orbitals with $2p$-orbitals on surrounding oxygen atoms; (ii) the bottom and the middle of the valence band is formed by the $\sigma$, $\pi$, and $\pi^*$ orbitals of the stretched O$_2$ molecules, hybridized with the $3p$-states of Si.   However, the most dramatic changes take place near the band gap of the system. We see that the $\sigma^*$ states of the oxygen pairs, surrounding a Si atom, no longer contribute to the conduction band of the system.  The spin-up $\sigma^*$ states become populated and are contributing now to the valence band maximum of the system, while the empty spin-down $\sigma^*$ states take a resonant position just above the valence band maximum in the band gap of the system. The situation is strikingly similar to the one of the polaron formation in Li$_2$O$_2$, where the extra electron populates the $\sigma$-antibonding state of the oxygen pair, leading to its significant elongation.\cite{Kang12} In our case the situation is very similar, but with an important exception: no extra charge is supplied to the system.   

Yet, the question remains: in the absence of electron injection, what is the source of the charge in the $\sigma^*$ orbitals? To delve further, we performed an analysis of charge transfer in Li$_2$O$_2$ upon Si substitution as shown in Fig.~\ref{drho}.   In this plot we present the difference between self-consistent charge density and the superposition of initial atomic charges, calculated in the plane, crossing one silicon atom, and a neighboring oxygen pair.  It can clearly be seen that charge is depleted on the Si site, and significant charge accumulation occurs on oxygen $\sigma^*$ antibonding orbitals, which is consistent with projected density of states analysis, presented in Fig. \ref{dos}b.  Overall, each elongated oxygen pair in an SiO$_6$ cluster has one electron more than a regular lithium peroxide O$_{2}^{2-}$ pair.   This additional charge is transfered from the Si atom, which serves in this system as ``additional source'' of electrons.         

\begin{figure}
 \includegraphics[width=8cm]{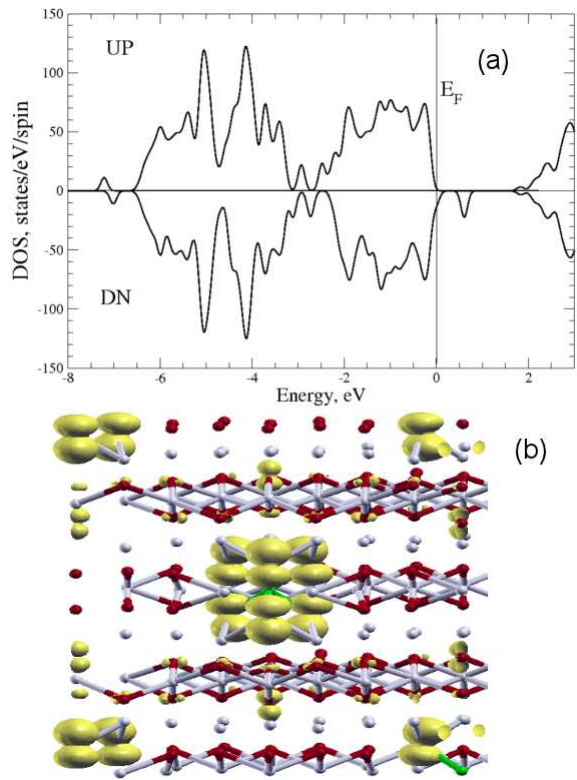}\\
  \caption{Panel (a) shows density of electronic states (DOS) for Si-doped Li$_2$O$_2$ 256-atom supercell, doped with one extra electron. Panel (b) shows a spacial charge distribution at the Fermi level for the system with one extra electron.}\label{charge}
\end{figure}

The most valuable effect of Si doping is, in our opinion, the induction of conducting states right above the valence band of the system. These states are formed by the $\sigma^*$ antibonding orbitals of the oxygen pairs, just as polaronic states are, and essentially preempt the formation of polarons. Meaning they should be stable upon electron injection, since the further deformation of oxygen pair is energetically quite costly. This is in contrast to pure Li$_2$O$_2$, where these states are located much higher in energy (above the band gap) and drop to form a polaron when an extra electron is injected. 

To verify this assumption, we performed 256-atom supercell calculations with one extra electron added to the system.  Fig.~\ref{charge}a shows the DOS for the charged system. It can be seen that the conducting $\sigma^*$ states are now partially occupied, and the DOS profile is only slightly modified, as compared to the undoped structure. We also computed the real-space distribution of electrons at the Fermi level of the doped system. Fig.~\ref{charge}b shows that the extra electron is shared between all oxygen pairs around Si impurities.   Lastly, to examine possible polaron formation somewhere outside the region of a Si impurity, we imposed a stretched O$_2^{2-}$ pair at the maximum possible distance from our impurity sites.  Upon structural relaxation, the charged system always prefered to return to initial state shown in Fig.~\ref{charge}b. 

To complete our study we provide a first order estimate of the electron mobility $\mu_e$ enhancement, that might be expected when electrons are injected into this doped system.  Mobility can be calculated using the Einstein relation\cite{Einstein56} 
$\mu_e=eD/k_B T$, where $k_B$ is Boltzmann's constant, and $T$ is a temperature. To first order the diffusion coefficient $D$ can estimated as $D=L^2/\tau$, where $L$ is the distance between $\sigma^*$ orbitals near neighboring Si atoms, and $\tau$ is the electronic state lifetime.   The inverse lifetime of electrons in a given $\sigma^*$-state can be estimated via: $1/\tau=\gamma/\hbar$, where $\gamma$ is the width of the $\sigma^*$ DOS peak, and $\hbar$ is Plank's constant.\cite{Datta05}
Substituting in the results from our current study, with a Si concentration of 1.6\% ($L\simeq7~\AA$, $\gamma=0.05~eV$), we obtained the room temperature mobility value of $\sim16~cm^2/[V s]$.  This value is significantly higher than polaronic hopping estimates calculated for pure lithium peroxide ($10^{-10}-10^{-9} cm^2/[V s]$).\cite{Kang12} The reason for this dramatic increase lies in the different charge transport mechanisms dominating the two systems: in stoichiometric Li$_2$O$_2$ charge carriers move through polaronic hopping, while in the proposed Si-doped structure charge transport likely occurs mainly through quantum mechanical tunneling between SiO$_6$ $\sigma^*$-states.   However, carrier transport assisted by these Si dopants deserves a careful detailed study on its own. We speculate that even a small increase in the Si concentration should lead to further increase of the spacial density of conducting $\sigma^*$-orbitals, leading to the formation of an intrinsic conducting network inside the Li$_2$O$_2$ crystal.   

%% \section{Conclusion}
In conclusion, we have presented a theoretical study of the electronic properties of Li$_2$O$_2$ crystal, where 1.6\% of lithium atoms were substituted by silicon. It has been demonstrated that under this type of doping additional conducting states are created in the band gap of the host crystal right above the top of the valence band. We have further demonstrated, that these conducting states originate from the antibonding states of the oxygen molecules, which have been partially populated by charge transfer from the silicon atoms. These results show that this charge transfer elongates the oxygen pairs, and makes them stable against polaron formation when an extra electron is injected in the system (effectively preempting polaron formation). The electron mobility of the proposed system is predicted to be significantly higher than that of intrinsic lithium peroxide, due to the non-polaronic nature of Si assisted carrier transport. Provided that each Si atom creates three elongated oxygen pairs, we believe that electronic transport the Li$_2$O$_2$ lattice should be significantly improved at reasonable concentrations of Si impurities.

\end{document}